\documentclass[12pt,letterpaper]{article}
\usepackage{amsmath}
\usepackage{amssymb}
\usepackage{latexsym}
\usepackage{hyperref}
\usepackage{enumitem}
\usepackage{graphicx}
\usepackage[export]{adjustbox}
\usepackage{natbib}
\newcommand{\be}{\begin{equation}}
\newcommand{\ee}{\end{equation}}
\newcommand{\bea}{\begin{eqnarray}}
\newcommand{\eea}{\end{eqnarray}}

\newcommand{\lag}{{\mathcal L}}

\newcommand{\uv}{{\mathrm{UV}}}
\newcommand{\ir}{{\mathrm{IR}}}
\newcommand{\eff}{{\mathrm{eff}}}

\begin{document} 

\title{The Quantum Field Theory on Which \\ the Everyday World Supervenes}

 \author{Sean M. Carroll\\
 Walter Burke Institute for Theoretical Physics\\
 California Institute of Technology, Pasadena, CA 91125\\
 and Santa Fe Institute, Santa Fe, NM 87501\\
 seancarroll@gmail.com}
\maketitle

\begin{quotation} \noindent
Effective Field Theory (EFT) is the successful paradigm underlying modern theoretical physics, including the ``Core Theory" of the Standard Model of particle physics plus Einstein's general relativity. I will argue that EFT grants us a unique insight: each EFT model comes with a built-in specification of its domain of applicability. Hence, once a model is tested within some domain (of energies and interaction strengths), we can be confident that it will continue to be accurate within that domain. Currently, the Core Theory has been tested in regimes that include all of the energy scales relevant to the physics of everyday life (biology, chemistry, technology, etc.). Therefore, we have reason to be confident that the laws of physics underlying the phenomena of everyday life are completely known.
\end{quotation}


\vfill
\noindent
Invited contribution to \textit{Levels of Reality: A Scientific and Metaphysical Investigation} (Jerusalem Studies in Philosophy and History of Science), eds. Orly Shenker, Meir Hemmo, Stavros Iannidis, and Gal Vishne. CALT 2021-005.

\newpage

\section{Introduction}

Objects in our everyday world -- people, planets, puppies -- are made up of atoms and molecules. 
Atoms and molecules, in turn, are made of elementary particles, interacting via a set of fundamental forces.
And these particles and forces are accurately described by the principles of quantum field theory.

We don't know whether relativistic quantum field theory is the right framework for a complete description of nature, and indeed there are indications (especially from black hole information and other aspects of quantum gravity) that it might not be.
But if we imagine describing nature in terms of multiple levels of reality, one such level appears to be a particular kind of quantum field theory, with other levels above (e.g. atoms and molecules; people and planets and puppies) and possibly other levels below.

In addition to a ``vertical'' division into levels, we can also consider carving each level ``horizontally" into different regimes, corresponding to different kinds of physical situations.
We might, for example, have a pretty good idea of how certain human beings will behave under ordinary conditions, but be less confident in how they will behave in extreme circumstances.
Within the domain of physics, we might distinguish between different regimes of energy or temperature or physical size.

In this paper I focus on the level of reality described by quantum field theory, in what we might call the ``everyday-life regime" (ELR) -- the energies, densities, temperatures, and other quantities characterizing phenomena that a typical human will experience in their normal lives.
This doesn't just mean, for example, the kinetic energy per particle that a human can muster under the power of their own musculature; it also includes phenomena such as sunlight that ultimately involve more extreme conditions in order to be explained.
It does not include conditions in the early universe, or near neutron stars or black holes, or involve phenomena such as dark matter and dark energy that don't interact noticeably with human beings under ordinary circumstances.

Modern physics has constructed an ``effective'' quantum field theory that purports to account for phenomena within this regime, a model that has been dubbed the ``Core Theory" \citep{wilczek2015beautiful}.
It includes the Standard Model of Particle Physics, but also gravitation as described by general relativity in the weak-field limit.
I will argue that we have good reason to believe that this model is both \emph{accurate} and \emph{complete} within the everyday-life regime; in other words, that the laws of physics underlying everyday life are, at one level of description, completely known.
This is not to claim that physics is nearly finished and that we are close to obtaining a Theory of Everything, but just that one particular level in one limited regime is now understood.
We will undoubtedly discover new particles and new forces, and perhaps even phenomena that are completely outside the domain of applicability of quantum field theory; but these will not require modifications of the Core Theory within the ELR, nor will the Core Theory fail to account for higher-level phenomena in that regime.
(A nontechnical version of this argument was given in \citep{carroll2017big}.)

 The interesting part of this claim is that it relies specifically on features of quantum field theory, which distinguish this paradigm from earlier models of physics.
 In particular, the effective field theory paradigm gives us good reason to believe that the dynamics of the known fields are completely understood, and the phenomenon known as ``crossing symmetry'' implies that any new particles or forces must interact too weakly with Core Theory fields to be relevant to everyday-life phenomena.
 In this paper I will explore this claim, starting with a precise statement of what the argument is supposed to be, and then a summary of the effective-field-theory approach. 
 I then discuss the specifics of the Core Theory, including why we are confident that its dynamics are understood in the ELR.
Then we will move to the feature of particle physics known as crossing symmetry, and how it constrains the possibility of unknown fields.
 I will then discuss the implications of these ideas for physics more broadly, and the wider project of understanding levels of reality.

\section{What Is Being Claimed}

\begin{figure}[h]
\centering
\includegraphics[width=.75\textwidth]{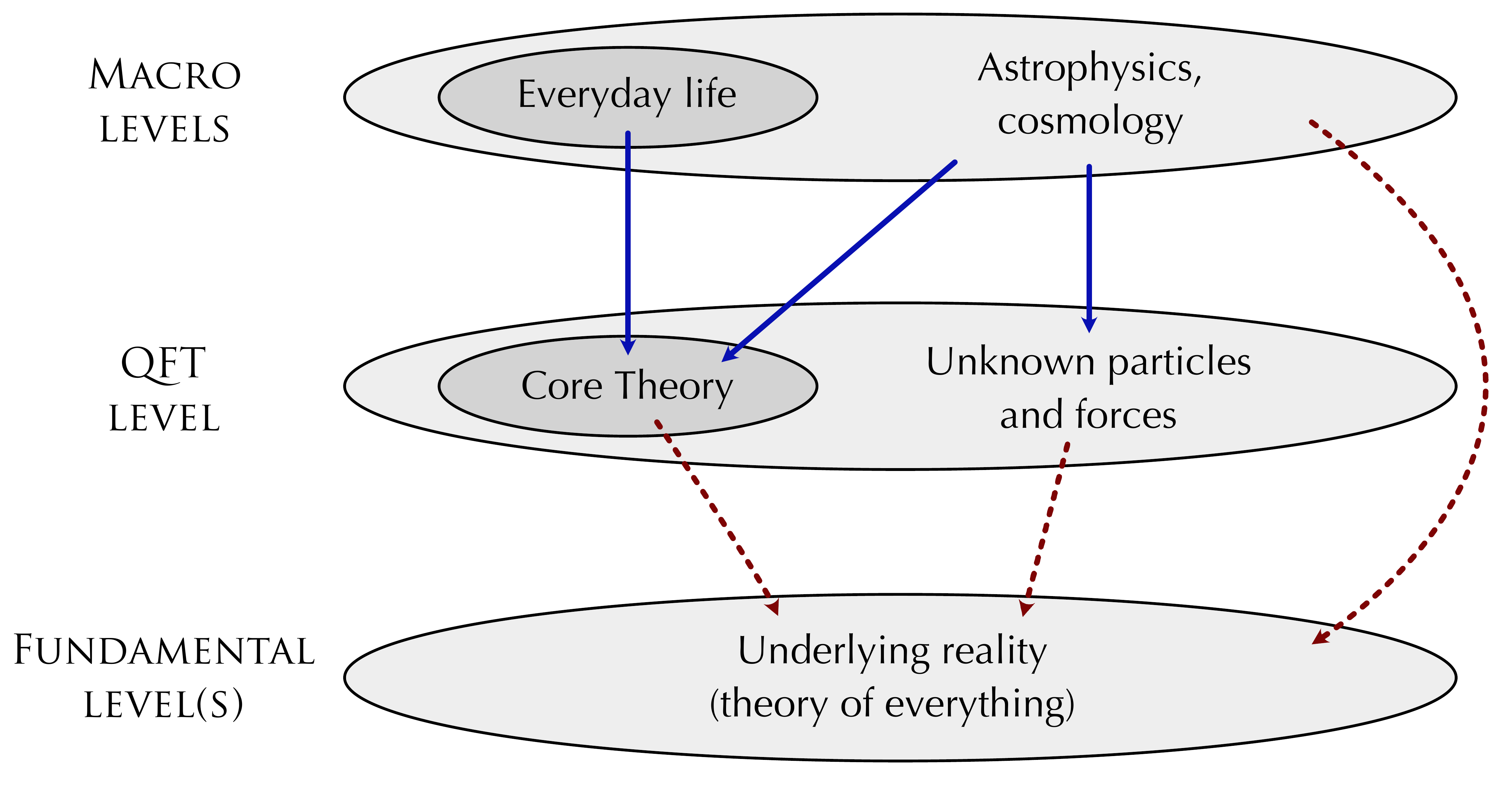}
\caption{\small Direct dependency relations between sets of phenomena at different levels. Solid blue arrows are established, while dashed red arrows are conjectural. Arrows that could be drawn, but are not, are relations we have good reason to think do not exist. So phenomena in the everyday life regime depend on the Core Theory, but not on unknown particles and forces, nor (directly) on an underlying theory of everything. Astrophysical phenomena depend on both the Core Theory and on new fields, and may depend directly on the underlying theory (e.g. in regimes where quantum gravity is important).}
\label{fig:dependence}
\end{figure}

The structure we are considering is portrayed in Figure~\ref{fig:dependence}, with levels of reality arranged vertically.
The middle ellipse is an effective relativistic quantum field theory, including weak-field quantum general relativity, thought of as a field theory on a flat background spacetime.
The smaller ellipse is the Core Theory of known particles and forces, with additional unknown particles and forces in the rest of the region.
The top ellipse summarizes all the more macroscopic levels, and is divided into the everyday-life regime (ELR) in the small ellipse, and more extreme astrophysical phenomena elsewhere.
(For our purposes here we can classify things like ultra-high-energy cosmic rays as astrophysical.)
Finally, we include a hypothetical level below, and therefore more fundamental than, effective quantum field theory.
I will refer to the theoretical explanations for what is described by each box as ``theories'' or ``descriptions'' or ``models," interchangeably.

The arrows in this figure indicate what phenomena depend on what other sets of phenomena; solid arrows are known relations, and dashed arrows are plausible but unknown.
The important claim being made is that certain arrows one could imagine drawing -- from ``Everyday life'' to ``Unknown particles and forces'' or ``Underlying reality" -- do \emph{not} appear.
In particular, everyday macro phenomena do not depend on either new particles/forces, nor directly on the underlying reality.
The Core Theory provides a complete and accurate description, we have good reason to believe, of everything on which macroscopic phenomena in the ELR supervene.
(In the next section we will be more specific about what is meant by the ELR.)

To make this claim more precise, let us distinguish between the Core Theory, which we know, and the idea of the Laws of Physics Underlying Everyday Life (LPUEL), whatever they might actually be.
We take it as established that everyday objects are at least partly made up of atoms, which are at least partly made of elementary particles, and that in some circumstances these particles interact through fundamental forces according to the standard understanding of physics, at least approximately.
The LPUEL, then, is whatever set of ingredients and dynamical rules operating at what we usually think of as the level of elementary particles that suffices to account for the properties of phenomena we experience in everyday life.
The Core Theory is a specific model, which we are arguing completely captures the LPUEL.
In principle, we might imagine a wide variety of ways in which the LPUEL deviate from the Core Theory; there might be heretofore undiscovered particles or forces that are relevant to the behavior of macroscopic phenomena, or quantum field theory itself might break down even within the ELR.
Our claim is that we have good reasons to believe this doesn't happen.

The argument will be as follows:
\begin{enumerate}
\item We have good reasons to believe that the LPUEL take the form of an effective quantum field theory (EQFT).
\item The Core Theory is an EQFT that to date is compatible with all known experimental data within the everyday-life regime.
\item Within the EQFT paradigm, the Core Theory could be modified in two possible ways: we could modify the dynamics of the known fields, or introduce additional fields.
\item Modified dynamics that could affect the LPUEL would require gross violations of the expectations of the EQFT paradigm, and are constrained experimentally.
\item Experimental constraints also imply that additional fields would be either too massive, too weakly-coupled, or too rare to affect the LPUEL.
\item Therefore, we have good reason to believe that the LPUEL are completely known.
\end{enumerate}

It's worth being especially careful about this claim, as it is adjacent to (but importantly different from) other claims that I do not support.
I am clearly not claiming that the correct theory of \emph{higher} levels is understood, which would be ludicrous.
Understanding atoms and particles doesn't help much with understanding psychology or economics.
I am not claiming that we understand all of particle physics; dark matter alone would be a persuasive counterexample.
Nor am I claiming that we are anywhere close to the end of physics, or achieving a theory of everything.
That may or may not be true, but is irrelevant to our considerations here; the correct theory of everything might require a relatively small extrapolation of our current understanding of quantum field theory, or it might ultimately involve a dramatically different and as-yet-unanticipated ontology that reduces to EQFT in some appropriate limit.
Regardless, the current claim is simply that the rules governing \emph{one} level of reality, in a particular circumscribed regime, are fully understood.
We don't know everything, and we don't know how close we are to knowing everything, but we know something, and we have a good understanding of the domain of applicability of that understanding.
Finally, I am not claiming any kind of ``proof'' that the Core Theory suffices, even when restricted to the ELR; as is always the case in science, all we can do is offer good reasons.

This argument goes somewhat beyond a simple assertion that a particular theory does a good job at explaining certain known phenomena.
The structure of quantum field theory allows us to predict the success of the model even in some circumstances where it has not yet been directly tested, given the basic assumptions on which QFT rests.
It is useful to contrast the situation with that of a theory such as Newtonian gravity.
The important rule there is the inverse-square law for the gravitational force,
\be
  \vec F = - \frac{GMm}{r^2} \hat{e}_r.
\ee
We might imagine testing this law, for example by comparing it with the motion of planets in the Solar System, and imagining that it might break down under circumstances in which it hasn't yet been tested.
Indeed, by now we know that it does break down for sufficiently large values of the gravitational potential $GM/r$, and corrections from Einstein's theory of general relativity become important, for example in computing the precession of the perihelion of Mercury.

But there was no way of knowing ahead of time what the domain of applicability of the theory was supposed to be, other than via direct experimental test.
It wasn't even possible to know what kind of phenomena would fall outside that domain.
It could be (and is) when the gravitational force was strong, but it also conceivably be when the force was extremely weak (and such theories have been suggested \citep{milgrom1983modification}).
Or when velocities were large, or when the angular momentum of the system pointed in certain directions, or when objects were made of matter rather than antimatter, or any number of other kinds of circumstances.

Quantum field theory is a somewhat different situation.
Any given EFT provides its own specification of what its domain of applicability will be (as we will specify in Section~\ref{sec:coretheory}), generally related to the energies and momenta characterizing particle interactions.
As long as the basic principles are respected (quantum mechanics, relativity, locality), we can be somewhat confident that our theory is accurate within this domain, even if we haven't tested it in some specific set of circumstances.
In that sense, we know a little bit more about the level of reality described by quantum field theory than we would have in other frameworks.

Our claim does have implications for how we should think about higher, emergent levels.
In particular, it highlights how very radical it is to imagine that understanding complex phenomena such as life or consciousness will require departures from the tenets of the Core Theory.
Such departures are conceivable, but we have good reasons to be skeptical of them.
The fact that the Core Theory is so robust and difficult to modify should count strongly against placing substantial credence in that kind of strategy.

\section{Effective Field Theory}

In this section I offer a brief review of quantum field theory and the Core Theory in particular.
It will necessarily be sketchy, but will serve to highlight the features that are relevant to our main point.
The notion of an effective field theory will be shown to place stringent constraints on the allowed dynamics of the known fields.

Quantum field theory is a subset of, rather than a successor to, quantum mechanics.
As in any quantum-mechanical theory, one has states represented by vectors in Hilbert space, an algebra of observables, and a Hamiltonian that evolves states forward in time.
In practice it is more common to work with a Lagrangian $L$ rather than a Hamiltonian; the Lagrangian is integrated over time to give an action $S$, which is exponentiated to provide a measure for a path integral.
In a ``local'' QFT, the Lagrangian can be written as a spatial integral of a Lagrange density $\lag$.
The Lagrange density, Lagrangian, and action are therefore related by
\be
  S = \int L\,dt = \int \lag \, d^4x,
  \label{action}
\ee
where $d^4x = dt\, d^3x$ is the volume element on spacetime, and in the path-integral formalism the amplitude for a transition between two specified configurations is
\be
A = \int [D\phi] e^{iS[\phi]}.
\label{pathintegral}
\ee
Here $\phi$ stands for all the degrees of freedom in the theory, $[D\phi]$ is a measure on the space of trajectories for those degrees of freedom, and we have suppressed an overall normalization factor.

We typically start with a classical Lagrange density -- most often referred to as simply the ``Lagrangian," with ``density'' taken as implied -- and then quantize it by one of various methods.
Given a set of fields, $\lag$ is some function of those fields and their spacetime derivatives.
It is often convenient to separate the terms appearing in $\lag$ into those that are quadratic in the fields, and those that are higher-order.
(Linear terms can be eliminated by re-defining fields so that such terms vanish in a stable vacuum state, while a constant term represents the vacuum energy, which we ignore in this discussion.)
The quadratic terms describe the ``free'' theory, and higher-order terms give interactions between the fields.

The free theory can be solved exactly in Fourier space, where the field is decomposed into modes of wave vector $\vec k$ and wave number $k=|\vec k|$, corresponding to wavelength $\lambda = 2\pi/k$.
These are associated with a momentum four-vector $p = (E/c,\vec p)$, where $\vec p = \hbar \vec k$.
(Henceforth we work in units where the speed of light $c$ and the reduced Planck constant $\hbar$ are set equal to one.)
For real particles, the energy satisfies $E^2={{\vec p}^2 + m^2}$, where $m$ is the mass of the field, but for virtual particles (interior lines in Feynman diagrams), $E$ is independent of ${\vec p}$.

In the free theory, the dynamics of any specific mode are that of a simple harmonic oscillator with frequency $E$.
Upon quantization, the quantum state can be represented as a superposition of discrete energy levels for each mode of every field.
These levels are interpreted as ``particles," which is how a quantum field theory can reproduce particle physics.
Fermionic fields give rise to matter particles such as leptons and quarks; bosonic fields give rise to forces, such as electromagnetism, the nuclear forces, and gravitation, as well as the Higgs field.
(We are obviously skipping a great many details, including the transformation properties of the fields under symmetry transformations.)

\begin{figure}[h]
\centering
\includegraphics[width=.3\textwidth, valign=c]{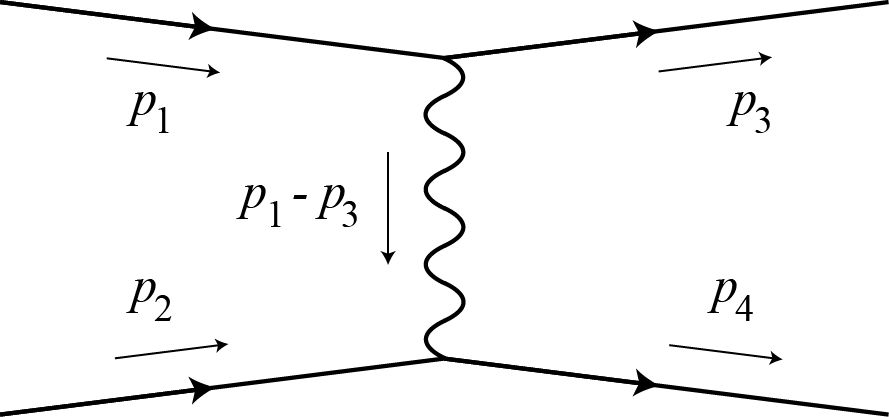}~~~~~~~~~
\includegraphics[width=.3\textwidth, valign=c]{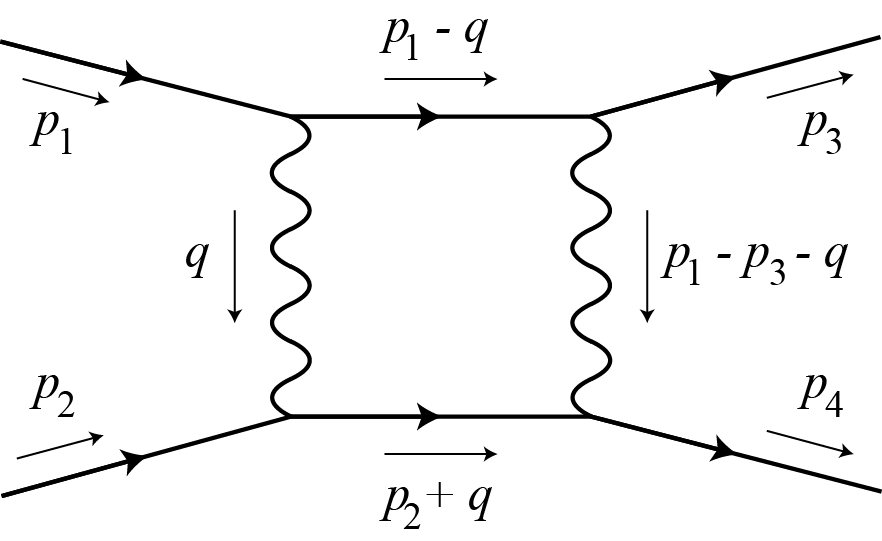}
\caption{\small Two Feynman diagrams for the scattering of two electrons (solid lines) by photons (waves).
In the tree diagram on the left, momentum conservation at each vertex fixes the momentum of the internal photon line; in the loop diagram on the right, a free momentum $q$ is integrated over.}
\label{fig:feynman}
\end{figure}

Feynman diagrams provide a convenient graphical way of representing particle interactions.
Lines entering from the left represent incoming particles, which interact by exchanging other particles, finally emerging on the right as outgoing particles.
Roughly speaking, classical effects are described by tree diagrams without any internal loops, while quantum corrections are described by loop diagrams.
The scattering amplitude for any specified process is obtained by adding the contributions from every possible diagram with the right incoming and outgoing particles.
Figure~\ref{fig:feynman} shows two contributions to the electromagnetic scattering of two electrons; first by the exchange of a single photon, and second by the exchange of two photons.

Each line in the Feynman diagram is labeled by the associated momentum four-vector.
Momentum is conserved at each vertex, so the sum of incoming momenta must equal the sum of outgoing momenta.
This condition suffices to fix the momenta of virtual particles (interior lines) in tree diagrams, but loop diagrams will have a number of undetermined momenta, one for each loop.
These loop momenta are integrated over to give the contribution of that diagram to the scattering amplitude.
The integration can include arbitrarily large momenta, and the resulting expressions often diverge, calling for some sort of renormalization procedure.
These high-momentum (short-wavelength) divergences are known as ``ultraviolet'' (UV) divergences, in contrast with infrared (IR) divergences from large numbers of massless particles in the incoming or outgoing states.

The modern attitude toward renormalization comes from the effective field theory program \citep{Manohar:2018aog,rivat2020philosophical}. 
This approach was systematized by Wilson \citep{Wilson:1971bg,Wilson:1971dh,Wilson:1973jj,Polchinski:1983gv}, though several of the important ideas had appeared earlier.
Divergences come from high-energy/short-wavelength virtual particles in loops.
But high energies and short wavelengths are precisely where we don't necessarily know the correct physical description.
High-mass particles that are irrelevant at low energies could be important in the UV, and for that matter spacetime and the entire idea of QFT might break down at small distances.
Fortunately, as Wilson emphasized, we don't need to understand the UV to accurately describe the IR. 
Let us introduce by hand an energy scale $\Lambda$, the ``ultraviolet cutoff."
The actual value of $\Lambda$ does not matter, as long as we consider incoming and outgoing momenta below that scale.
In practice the effect of the cutoff is that we only integrate the momenta of virtual particles in loops up to the value of $\Lambda$, rather than all the way to infinity.
This renders the loop integrals finite, though they do depend on $\Lambda$.

The physical predictions of the theory itself, however, do not depend on $\Lambda$.
Rather, the original action defining the theory is replaced by an effective action $S_{\eff}$ for the IR modes alone.
Schematically, from the path-integral perspective we have
\bea
A &=& \int [D\phi] e^{iS[\phi]}\\
&=& \int[D\phi_\ir][D\phi_\uv] e^{iS[\phi_\ir, \phi_\uv]}\\
&=& \int [D\phi_\ir]e^{iS_{\eff}[\phi_\ir, \Lambda]}, \label{effectivepathintegral}
\eea
where $\phi_\uv$ represents UV modes (momenta greater than $\Lambda$) and $\phi_\ir$ represents IR modes (momenta less than $\Lambda$).

Crucially, \emph{the effective action will describe the dynamics of a local quantum field theory}, even though we have integrated out some of the degrees of freedom.
Roughly speaking this is because we have eliminated modes with wavelengths less than $\Lambda^{-1}$, while considering only the dynamics of particles than can probe length scales greater than $\Lambda^{-1}$.
The effective action $S_\eff$ is the integral of an effective Lagrangian $\lag_\eff$, which can be written as a power series in the field operators.
It will generally include an infinite number of terms, with arbitrarily high powers of the fields.
The higher-order terms will be parameterized by coefficients that depend on the cutoff $\Lambda$, in such a way that all of the dependence on $\Lambda$ completely cancels in any physical process for purely IR particles.
Predictions of the effective field theory are thus independent of the arbitrary cutoff.

In presenting things this way, we have spoken as if the fundamental QFT is valid to all energies, even if we are only considering an effective theory of the IR modes.
Whether or not that is the case, quantum field theory still seems to be the universal form that physical theories take in the low-energy limit, given certain assumptions.
This phenomenon of ``universality'' means that the most fundamental theory might feature superstrings, or discrete spacetime, or some more dramatic departure from the relativistic QFT paradigm, and still look like an EFT at low energies.
 \citet{weinberg1995quantum} has argued that the following assumptions suffice:
\begin{itemize}
\item Quantum mechanics.
\item Lorentz invariance.
\item Cluster decomposition.
\item The theory describes particle-like excitations at low energies.
\end{itemize}
(Cluster decomposition is a kind of locality requirement, that amplitudes for widely-separated scattering events be independent of each other.)
This is not a rigorous result, but what \citet{Weinberg:1996kw} refers to as a ``folk theorem." 
Nevertheless, it is consistent with everything we know about the universality of QFT from a variety of ``ultraviolet completions," which themselves may or may not be QFTs.
The explicit arguments for it only hold in the perturbative regime where fields are relatively small deviations away from the vacuum; hence, it fails to apply to strong-field phenomena like black holes.

None of these listed assumptions is inviolate.
Quantum mechanics could be incomplete, and Lorentz invariance or locality could be merely approximate.
Nevertheless, they have been tested to impressive accuracy in experiments.
Without favoring any particular stance toward the correct theory of everything describing reality might be, it makes sense to believe that the world follows the rules of effective field theory in the long-distance/low-energy perturbative regime.

These considerations are enough to eliminate one particular dependency relation that we could imagine drawing in Figure~\ref{fig:dependence}: from everyday macro phenomena directly down to underlying reality, bypassing the QFT level.
In other words, to the extent that we have good reasons to believe that the low-energy behavior of reality is accurately modeled by an effective quantum field theory, and that everyday phenomena are within that regime, we have good reason to think that there are no non-QFT phenomena characteristic of the theory of everything that are relevant for the everyday-life regime.

\section{The Core Theory}
\label{sec:coretheory}

We know more than just the general claim that low-energy physics is described by an effective quantum field theory; we know what theory it is.
The Core Theory is an effective field theory that contains the well-known Standard Model of particle physics, but also quantum general relativity in the weak-field limit.
The lack of a full theory of quantum gravity is a well-known outstanding issue in theoretical physics, but we have a perfectly adequate \emph{effective} theory of quantum gravity in this regime.
``Weak-field" here means essentially ``small Newtonian gravitational potential $GM/r$," which includes everything we observe other than black holes, the very early universe, and perhaps neutron stars. 
It certainly covers planets in the Solar System and apples falling from trees (and for that matter gravitational waves).

In path-integral form, the theory is given by
\bea
&A = \displaystyle\int_{k < \Lambda} [Dg][DA][D\psi][D\Phi]\exp\bigg\{i \int d^4x\,\sqrt{-g}\bigg[\frac{1}{16\pi G}R - \frac{1}{4}F_{\mu\nu}F^{\mu\nu} 
+ i\bar\psi\gamma^\mu D_\mu \psi 
 \nonumber  \\&
+ |D_\mu\Phi|^2 -V(\Phi) + \left(\bar\psi^i_L Y_{ij}\Phi\psi^j_R + \mathrm{h.c.}\right) + \sum_a \mathcal{O}^{(a)}(\Lambda) \bigg]\bigg\}.
\label{coretheory}
\eea
This is of the general form (\ref{effectivepathintegral}), with an action given by a spacetime integral as in (\ref{action}).
Specific terms in the Lagrange density (large square brackets) include $R$ for gravity, $F_{\mu\nu}F^{\mu\nu}$ for the gauge fields of the strong, weak, and electromagnetic interactions, $\bar\psi \gamma D \psi$ for the kinetic energy of the fermion fields, $|D\Phi|^2$ for the kinetic energy of the Higgs, $V(\Phi)$ for the Higgs potential, and $\bar\psi Y\Phi\psi$ for the Higgs-fermion interaction.
(Interactions between gauge fields and fermions are hidden in the gauge-covariant derivative $D_\mu$, and interactions between gravity and other fields are both there and in the overall volume element $\sqrt{-g}$ outside the brackets.)
Details can be found in standard QFT texts \citep{peskin2015introduction}.
A crucial role here is played by the notation $k<\Lambda$ in the overall path integral, a reminder that this is an effective theory only applicable for momenta below the cutoff.
The term $\sum\mathcal{O}^{(a)}(\Lambda)$ represents an infinite series of higher-order terms, each of which depend on (and in general will be suppressed by powers of) the cutoff.
These terms ensure that physical predictions are independent of the cutoff value.

This is the theory that seems to underlie the phenomena of our everyday experience.
The Higgs field gets an expectation value in the vacuum, breaking symmetries and giving masses to fermions.
Quarks and gluons are confined into bound states such as nucleons and mesons.
At low temperatures, most heavy particles decay away, leaving only protons, neutrons, electrons, photons, neutrinos, and gravitons, the latter two of which interact so weakly as to be essentially irrelevant for everyday phenomena.
(Classical gravitational fields, which are relevant, can be thought of arising from virtual gravitons, but individual real gravitons are not.)
Protons and neutrons combine into nuclei, which capture electrons electromagnetically to form atoms.
A residual electromagnetic force between atoms creates molecules, and underlies all of chemistry.
Finally, all of the resulting objects attract each other via gravity.
Aside from nuclear reactions, everyday objects are made of electrons and roughly 254 species of stable nuclear isotopes, interacting through electromagnetic and gravitational forces.

What value for $\Lambda$ should we choose?
Low-energy predictions are independent of the specific value of $\Lambda$, as long as we choose it to be higher than the characteristic momentum scales of whatever processes we would like to consider.
But it should also be lower than any scale at which potentially unknown physics could kick in (massive particles, restored symmetries, discrete spacetime, etc.).
In practice, this means we should take $\Lambda$ to be no higher than scales we have probed experimentally.
For the Core Theory, we should be able to safely put the cutoff at least as high as 
\be
  \Lambda_\mathrm{CT} = 10^{11}~\mathrm{electron~volts~(eV)},
  \label{LambdaCT}
\ee 
a scale that has been thoroughly investigated at particle accelerators such as the Large Hadron Collider.
(Proton-proton collisions at the LHC have a center-of-mass energy of $10^{13}$\,eV, but that is distributed among a large number of particles; $10^{11}$\,eV is a reasonable value for the energy up to which individual particle collisions have been explored.)
Much above that scale, and new physics is possible, and indeed many physicists are still hopeful to find evidence for supersymmetry, large extra dimensions, or other interesting phenomena.

Let us compare this to the everyday-life regime (ELR), which we are finally in position to define more precisely.
The domain of applicability of an EFT is characterized by energy -- more precisely, by the relative momenta of interacting particles as measured in their overall rest frame.
If these momenta are all below the cutoff scale $\Lambda$, the model should be accurate.
(Note that the relevant quantity is the energy per particle, not the total energy of an object, which for macroscopic objects can be quite large.)
In the everyday macroscopic world, typical energies of interest are those of chemical reactions, typically amounting to a few electron volts (eV).
The binding energy of an electron in a hydrogen atom is 13.6\,eV, while the bond between two carbon atoms is 3.6\,eV.
Bulk macroscopic motions are typically well below this energy scale; the kinetic energy of a proton in a speeding bullet is about 0.01\,eV. 

We might want to include nuclear reactions, such as occur in the interior of the Sun. 
The relevant energies are $10^{8}$\,eV or below; for example, the fusion reaction converting deuterium and tritium into helium plus a neutron releases $1.8\times 10^{7}$\,eV of energy.
An expansive definition of the ELR, building in a bit of a safety buffer, might therefore include interactions at or below an energy of 
\be
  E_\mathrm{ELR} = 10^9\,\mathrm{eV}.
  \label{ELR} 
\ee
All of the interactions of the particles and forces around us, and all of the radiation we absorb and admit, occurs at energies per particle lower than this value (unless we are hanging out at a high-energy particle accelerator).

The fact that $E_\mathrm{ELR} < \Lambda_\mathrm{CT}$ implies that the domain of applicability of the Core Theory encompasses the everyday-life regime.
This seems to imply that not only can we list the quantum fields out of which everyday phenomena are made, but we know what their dynamics are.
One loophole comes from the existence of the infinite series of higher-order terms $\sum\mathcal{O}(\Lambda)$ that inevitably appear in an effective Lagrangian.
Should we be confident that they don't affect the dynamics in important ways, even at low energies?

We can gain insight by simple dimensional analysis.
With $\hbar=c=1$, energy and mass have the same units, which are the same as the units of inverse length and inverse time, and the Lagrange density has units of energy to the fourth power.
Consider a real scalar field $\phi$ with units of energy.
The part of its effective Lagrangian that contains only that field (no other fields or spacetime derivatives) is the potential energy, which takes the form
\be
  V_\eff(\phi) = \frac{1}{2}m^2\phi^2 + c_3\Lambda \phi^3 + c_4 \phi^4 + \frac{c_5}{\Lambda}\phi^5 + \frac{c_6}{\Lambda^2}\phi^6 + \cdots.
\ee
Here $m$ is the (renormalized) mass of the field, the $c_i$s are dimensionless coefficients, and appropriate powers of the cutoff $\Lambda$ appear to ensure that each term has units of (energy)$^4$.

The specific values of the $c_i$s will depend on $\Lambda$ (the phenomenon known as renormalization group flow), in such a way as to render physical predictions independent of $\Lambda$.
But we have a ``natural'' expectation that these dimensionless parameters should be of order unity, rather than extremely large or small.
It would be interesting to interrogate this notion of naturalness in a philosophically rigorous way, but for now we will merely note that this is indeed what happens in explicit models of EFTs where the complete UV completion is known and the parameters can be calculated as a function of $\Lambda$.

The terms in $\lag_\eff$ can be characterized as ``relevant'' if they appear with positive powers of $\Lambda$ (or other quantities with dimensions of energy, like $m$), ``marginal'' if they are of order $\Lambda^0$, and ``irrelevant'' if they appear with negative powers of $\Lambda$.
This reflects the fact that for energies well below $\Lambda$, terms with negative powers of $\Lambda$ become increasingly irrelevant for making predictions.
(It is these terms that are classified as ``non-renormalizable.")
But we've already said that our EFT is meant to be applicable only for momenta well below $\Lambda$.
Therefore, our strong expectation is that these higher-order terms are indeed irrelevant for the dynamics of Core Theory fields in the ELR.
(For explicit experimental constraints see \citet{burgess1994model}.)
The action we wrote for the Core Theory already includes all of the relevant and marginal terms that are consistent with the symmetries.
We not only know what the basic fields are, but we have good reason to think that we know how they behave to very high accuracy.

\section{New Particles and Forces}

If we believe we understand the dynamics of the known fields of the Core Theory, the other way that model could fail to completely account for everyday phenomena -- without leaving the EFT paradigm entirely -- is if there are unknown fields that could play a subtle but important role.
We can distinguish between three ways this could happen.
\begin{itemize}
\item A new field could show up as virtual particles mediating a new kind of interaction between the known fields.
However, this would essentially modify the low-energy effective action (\ref{coretheory}) of the Core Theory.
This would have no observable effects unless the results deviated significantly from our effective-field-theory expectations, and as we have noted there are good constraints on any such possibility.
So we will not consider this alternative in detail.
\item A field could give rise to new long-lived particles that played a distinct dynamical role in macroscopic phenomena, much like electrons, protons, and neutrons do.
Such a particle could be ambient in the universe, much like dark matter but possibly with a lower overall energy density.
Perhaps a particle of this form participates in the neurochemical processes of conscious creatures \citep{pullman2000}.
\item A weakly-interacting bosonic field could condense to give a classical force field, what physicists think of as a ``fifth force."
Such a force could conceivably induce interactions between neurons, or even between different brains, as two vivid examples.
\end{itemize}
Let's consider these last two possibilities in turn.

In contemplating the existence of novel ambient particles, it is useful to compare with the case of neutrinos, which are known to exist.
There are a lot of neutrinos in the universe; the flux near Earth, from both the cosmic neutrino background and solar-generated neutrinos, is of order 10 trillion neutrinos per square centimeter per second.
But they interact with ordinary matter quite weakly (literally through the ``weak interactions'' of the Standard Model), so much so that of the order $10^{21}$ neutrinos that pass through a typical human body in a typical lifetime, approximately one of them will actually interact with the atoms in that body.
Any hypothetical new particle would have to have substantially higher interaction strength with ordinary matter in order to play a role in everyday phenomena.

\begin{figure}[h]
\centering
\includegraphics[width=.80\textwidth, valign=c]{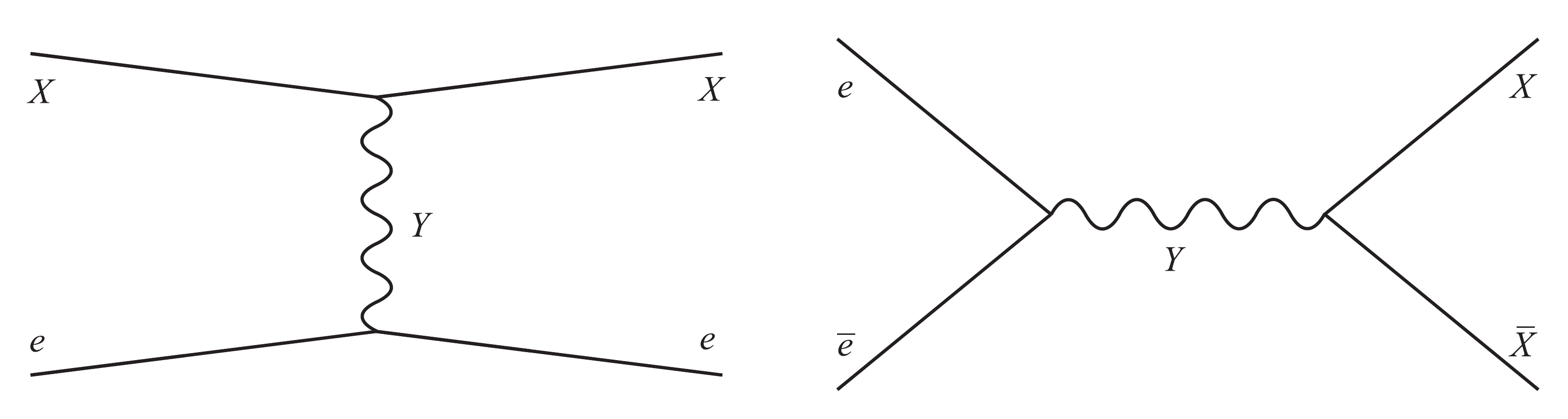}
\caption{\small Crossing symmetry relates the amplitudes for these two processes, an interaction of a new particle $X$ with an electron $e$ via a mediator $Y$, and annihilation of an electron/positron pair into an $X$/anti-$X$. Any new particle that interacts with ordinary matter can therefore be created in particle collisions.}
\label{fig:crossing}
\end{figure}

One way of constraining such new particles is by simply trying to create them at particle accelerators.
The QFT property of crossing symmetry guarantees that such searches are feasible.
Consider a new particle $X$ that interacts with electrons through some new force, mediated by a new field $Y$; something along these lines would be necessary for $X$ to affect everyday objects.
In Feynman-diagram language we can represent that as an incoming electron and $X$, which interact via virtual $Y$ exchange and then continue on.
Crossing symmetry implies that the amplitude for such an interaction will be related to that obtained by rotating the diagram by ninety degrees, and interpreting particles going backward in time as antiparticles.
Hence, this scattering amplitude is related to the amplitude for an electron and positron (anti-electron) to annihilate into a $Y$, which then decays to an $X$ and an anti-$X$, as shown in Figure~\ref{fig:crossing}.

Fortunately, colliding particles together and studying what comes out is particle physicists' stock in trade.
Our $X$ particle must be electrically neutral and invisible to the strong nuclear force, otherwise it would interact very noticeably and have been detected long ago.
It therefore won't lead a visible track in a particle detector, but there are indirect methods for constraining its existence.
For example, new particles give other particles new ways to decay, decreasing their lifetime and therefore increasing the width of energy distribution of particles into which they decay.
(This can be thought of as a consequence of the energy-time uncertainty principle; faster decay implies more uncertainty in energy.)
The decay width of the $Z$ boson was measured to high precision by the Large Electron-Positron Collider, a predecessor to the Large Hadron Collider at CERN.
Results are usually quoted in terms of the number of ``effective neutrino species," although the principle applies to non-neutrino particles as well.
(Even if $X$ coupled to quarks and not to electrons, it would still be produced by interactions with virtual quarks.)
There are three conventional neutrino species in the Core Theory, and the LEP measurement came in at $2.9840 \pm 0.0082$ \citep{mele2015measurement}.
We can interpret this as saying that there are no unknown particles with masses less than half that of the $Z$ (about $4\times 10^{10}$\,eV) that interact with Core Theory fermions with an interaction strength greater than or equal to that of neutrinos.\footnote{One subtlety is that the electron-$X$ interaction could be enhanced if the two particles exchanged a large number of virtual $Y$s; something similar happens in ordinary electromagnetism. But that would require the $Y$ itself to be a very light particle, and then it would contribute the number of effective neutrino species bounded by LEP.}

Heavier $X$ particles can also be constrained, and other measurements also provide limits \citep{Acciarri_1999,Fox_2012,Aad:2020cws}.
If $X$ particles are extremely heavy, say over $10^{11}$\,eV, they would be out of reach of current particle accelerators. 
But if such particles are ambient, there is a limit on how abundant they can be, given by the dark-matter density. 
(If new stable particles have more mass density than dark matter, they would be ruled out by astrophysical measurements.)
So as not to have more mass density than dark matter, an ambient particle of mass $m$ must have a number density lower than about $(3\times 10^{11}$\,eV$/m)$ per liter in the Solar System. It is hard to imagine such dilute particles being relevant for everyday dynamics.

We have noted that neutrinos barely interact with ordinary matter at all; any hypothetical new ambient particle that would be relevant to the behavior of macroscopic objects would have to interact much more strongly than that.
Particle-physics constraints imply that there are no such particles. 
New particles may certainly exist, but they must be either short-lived, weakly-interacting, or extremely rare in the universe.
We can therefore conclude that unknown ambient particles do not play a role in accounting for phenomena in the everyday-life regime.

The other reasonable option is the existence of a bosonic field that couples weakly to individual particles, so that direct searches for the boson would be fruitless, but that is sufficiently low-mass that it can accumulate to give rise to a macroscopic force field.
(The range of a field is inversely proportional to its mass, with $r \mathrm{[cm]} \sim 2\times 10^{-5}/(m$\,[eV]).)
For our purposes here we could define ``macroscopic" as larger than one micrometer; the average cell in a human body is between 10 and 100 micrometers in diameter.

Gravity itself is an example of a field whose quanta are undetectable but that gives rise to a macroscopic force.
Individual gravitons couple far too weakly to be detected, but the net gravitational force sourced by matter in the Earth is enough to keep us anchored to the ground, because the gravitational field is infinite-range (gravitons are massless) and every particle contributes positively to the force.
Gravity is nevertheless extremely weak; the gravitational force between two typical human bodies separated by a distance $d$ is less than $10^{-7}$ the electromagnetic force between two individual protons at the same separation.
To be generated by human-sized (or smaller) objects, and yet have a noticeable impact on the dynamics of the macroscopic world, a new force would have to be enormously stronger than gravity.
This seems unlikely at first glance, as we would presumably have noticed such a force.
But it's conceivable that it couples only to certain combinations of particles (rather to everything, as gravity does), and that it has a macroscopic but finite range, so that it doesn't affect celestial dynamics or apples falling from trees.
It's therefore worth examining the possibility more carefully.

\begin{figure}[h]
\centering
\includegraphics[width=.8\textwidth, valign=c]{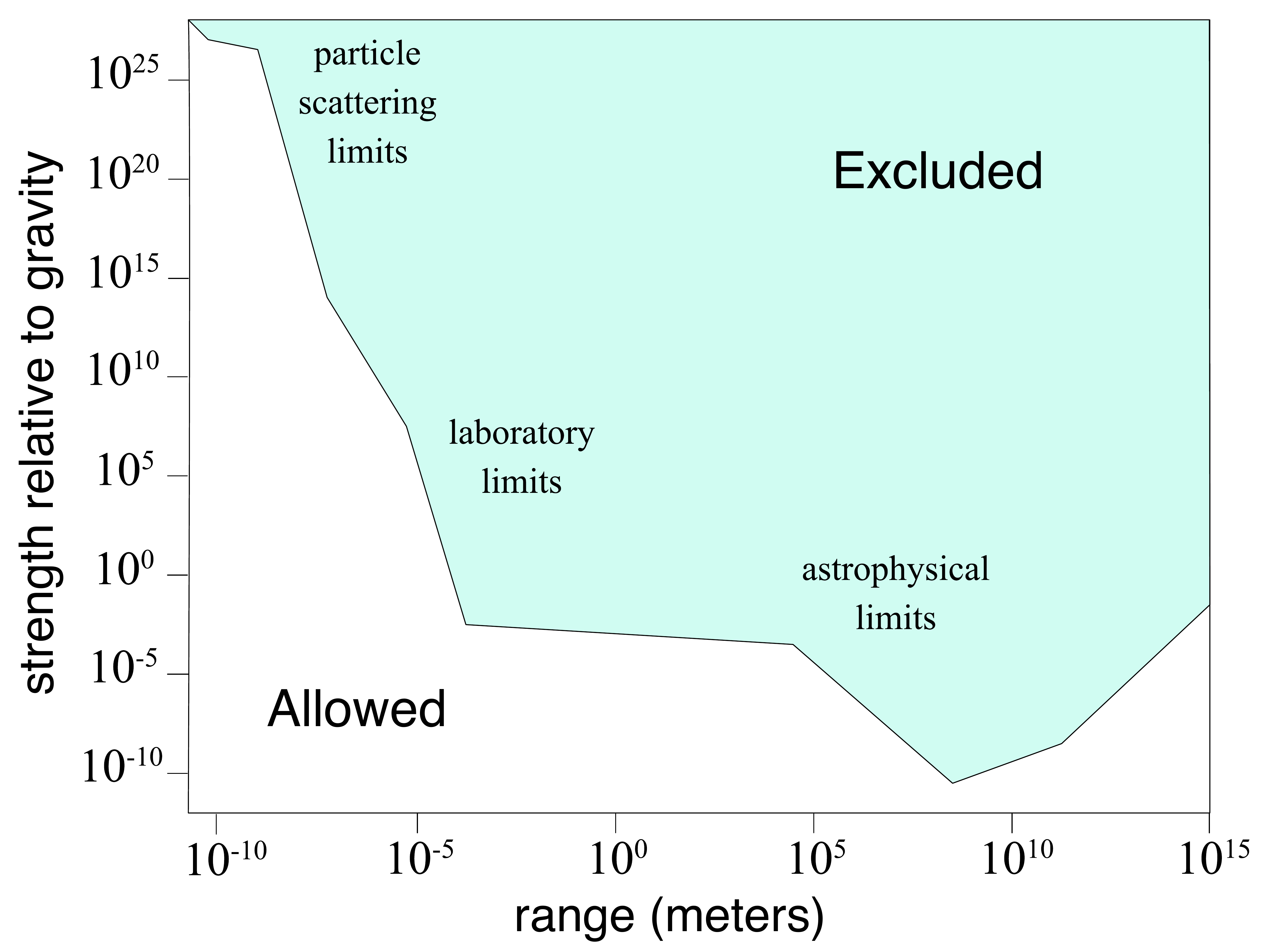}
\caption{\small Limits on a new fifth force, in terms of its strength relative to gravity, as a function of its range. 
Adapted from data collected in \citep{ADELBERGER2009102}.
This is a rough reconstruction; see original source for details.}
\label{fig:fifth-force}
\end{figure}

Fortunately, there aren't that many different ways in which a fifth force can couple to ordinary matter.
Within the framework of low-energy effective field theory, we can think of the source of the new force as some linear combination of electrons, protons, and neutrons.
The available parameter space can be constrained by measuring the forces between macroscopic objects of substantially different chemical compositions.
We don't need to be too precise about the results here, as a rough guide is more than adequate for our purposes.
From a variety of experimental and astrophysical techniques, stringent bounds have been placed on the possible existence of new long-range forces \citep{ADELBERGER2009102}; the results are summarized in Figure~\ref{fig:fifth-force}.

It is clear from examination of this plot that for ranges greater than $10^{-4}$\,m (100 micrometers), any new force must be weaker than gravity, and at $10^{-3}$\,m and above the limits are better than $10^{-3}$ gravity.
Given how weak gravity itself is between human-sized objects, this definitively rules out the possibility that such forces are important for dynamics in the ELR.
At shorter ranges the limits deteriorate, both because the magnitude of the force between small test objects is smaller and harder to measure, and (more importantly) because it becomes harder to eliminate possible contamination from residual electromagnetic forces.
For precisely this reason, such forces will also be irrelevant for macroscopic dynamics.
At one micrometer, a force $10^9$ times gravity would be allowed, but that is only $10^{-27}$ times the strength of electromagnetism.
Even with substantial cancellations between positive and negative charges, residual electromagnetic forces will overwhelm a fifth force at these ranges.
All the way down at atomic scales, $\sim 10^{-10}$\,m, any new force must still be less than $10^{-6}$ the strength of electromagnetism.

We therefore conclude that, within the framework of effective field theory, there is no room for unknown fields or unanticipated dynamics to play a role in accounting for macroscopic phenomena in the everyday-life regime.
There can be, and very likely are, more fields yet to be discovered, but they must either be extremely dilute in the universe so that we essentially never interact with them, or so weakly coupled to ordinary matter that they exert essentially no influence.
Quantum field theory might not, and probably is not, the correct framework in which to formulate an ultimate theory of everything, but given certain plausible assumptions low-energy physics will nevertheless be accurately modeled by an EFT, so everyday phenomena do not depend directly on deeper levels, only through the Core Theory.
There is much of physics that we don't know, and it is entirely unclear how close we are to achieving a fundamental theory of nature.
But we do understand the laws of physics underlying everyday phenomena as described at one particular level of reality, that of effective quantum field theory.

\section{Discussion}

I have argued that we have good reasons to believe that everyday-life phenomena supervene on the Core Theory, and not on as-yet-undiscovered particles and forces or on new principles at more fundamental levels.
The argument relies on an assumption that the world is entirely physical, and that there is a level of reality accurately described by an effective quantum field theory.
Then the general properties of quantum field theory, plus known experimental constraints, lead us to the conclusion that the Core Theory suffices.

If this package of claims -- physicalism, EFT, Core Theory -- is correct, it has a number of immediate implications.
There is no life after death, as the information in a person's mind is encoded in the physical configuration of atoms in their body, and there is no physical mechanism for that information to be carried away after death.
The location of planets and stars on the day of your birth has no effect on who you become later in life, as there are no relevant forces that can extend over astrophysical distances.
And the problems of consciousness, whether ``easy'' or ``hard,'' must ultimately be answered in terms of processes that are compatible with this underlying theory.

Less obviously, our understanding of the Core Theory has implications for the development of technology.
Historically, progress in fundamental physics (as it was defined at the time) has often had important technological implications, from mechanics and electromagnetism to quantum theory and nuclear physics.
That relationship has largely evaporated.
The last advance in fundamental physics (defined in a modern context as new particles or forces or dynamics at the quantum-field level) to be put to use in technology was arguably the discovery of the pion in 1947.
Since then, technological development has depended on increasingly sophisticated ways of manipulating the known particles and forces in the Core Theory.
This is likely to be the case for the foreseeable future; the kinds of new particles remaining to be discovered either require multi-billion-dollar particle accelerators to produce (and even then they decay away in zeptoseconds), or they interact with ordinary matter so weakly as to be essentially impossible to manipulate in useful ways.
It is hard to imagine technological applications of such discoveries.
Even quantum computing, which has involved important conceptual breakthroughs, makes use of the same underlying physical matter and laws that have been known for well over half a century.

Needless to say, the claim that we fully understand the laws of physics underlying everyday life might very well be incorrect, even if there are good reasons to accept it.
It is easy enough to list some potential loopholes to the argument, ways in which the claim might fail to be true by going outside the EFT framework.
\begin{itemize}
\item Violations of locality.
In the context of an EFT, locality of interactions implies that the electromagnetic or gravitational fields (or unknown fifth-force fields) produced by an object are simply the net fields produced by each of the constituent particles individually.
Outside the EFT paradigm, we could imagine forces that depend non-locally on sources, so that whether or not a force is produced would depend on the specific arrangement of particles within it.
Such a force might not be produced by a collection of electrons, protons, and neutrons in the form of a cantaloupe, for example, but be produced by the same particles when they are in the form of a human brain.
To the best of my knowledge, this possibility has not been investigated carefully (and to be honest, there is not a lot of motivation for it).
\item Quantum wave function collapse.
In conventional quantum mechanics, the probability of a measurement outcome is given by the absolute-value squared of the corresponding amplitude of the wave function (the Born Rule).
Other than that, the process is thought to be entirely random, with no structure other than that statistical rule.
But perhaps it is not, and quantum systems evolve in subtle and specific ways to bring about particular outcomes.
This scenario has been studied, typically in the context of trying to attain a better understanding of consciousness \citep{penrose1989emperor,Chalmers2014ConsciousnessAT}.
\item Departures from physicalism.
Everything we have said presumes from the start that the world is ultimately physical, consisting of some kind of physical stuff obeying physical laws.
There is a long tradition of presuming otherwise, and if so, all bets are off.
The well-known issue is then how non-physical substances or properties could interact with the physical stuff.
\end{itemize}
This list is not meant to be exhaustive, but provides a flavor of the options available to us.

The reasons for denying the claim advanced in this paper, and going for one of the above loopholes instead, generally arise from a concern that the physical dynamics of the Core Theory cannot suffice to account for higher-level phenomena, whether the phenomenon in question is life after death or the experience of qualia.
Our considerations do not amount to an airtight proof (which would be essentially impossible), but they do highlight the challenge faced by those who think something beyond the Core Theory is required.
The dynamics summarized in equation (\ref{coretheory}) are well-defined, quantitative, and unyielding, not to mention experimentally tested to exquisite precision in a wide variety of contexts.
Given a quantum state of the relevant fields, it accurately predicts how that state will evolve.
Skeptics of the claim defended here have the burden of specifying precisely how that equation is to be modified.
This would necessarily raise a host of tricky issues, such as conservation of energy and unitary evolution of the wave function.
A simpler -- though still extremely challenging -- alternative is to work to understand how those dynamics give rise to the emergent levels of reality in our macroscopic world.

\section*{Acknowledgements}
 
It is a pleasure to thank Jenann Ismael, Ira Rothstein, Charles Sebens, and Mark Wise for helpful comments on a draft version of this manuscript.
This research is funded in part by the Walter Burke Institute for Theoretical Physics at Caltech, by the U.S. Department of Energy, Office of Science, Office of High Energy Physics, under Award Number DE-SC0011632, and by the Foundational Questions Institute.

\bibliographystyle{apa-good}

\bibliography{qft-el}

\begin{thebibliography}{21}
\expandafter\ifx\csname natexlab\endcsname\relax\def\natexlab#1{#1}\fi
\expandafter\ifx\csname url\endcsname\relax
  \def\url#1{{\tt #1}}\fi
\expandafter\ifx\csname urlprefix\endcsname\relax\def\urlprefix{URL }\fi

\bibitem[{Aad et~al.(2020)}]{Aad:2020cws}
Aad, G., et~al. (2020).
\newblock {Dijet resonance search with weak supervision using $\sqrt{s}=13$ TeV
  $pp$ collisions in the ATLAS detector}.
\newblock {\em Phys. Rev. Lett.\/}, {\em 125\/}(13), 131801.

\bibitem[{Acciarri(1999)}]{Acciarri_1999}
Acciarri, M. e.~a. (1999).
\newblock Single and multi-photon events with missing energy in $e$-$\bar e$
  collisions at $\sqrt{s}$ = 189 gev.
\newblock {\em Physics Letters B\/}, {\em 470\/}(1-4), 268–280.
\newline\urlprefix\url{http://dx.doi.org/10.1016/S0370-2693(99)01286-1}

\bibitem[{Adelberger et~al.(2009)Adelberger, Gundlach, Heckel, Hoedl, \&
  Schlamminger}]{ADELBERGER2009102}
Adelberger, E., Gundlach, J., Heckel, B., Hoedl, S., \& Schlamminger, S.
  (2009).
\newblock Torsion balance experiments: A low-energy frontier of particle
  physics.
\newblock {\em Progress in Particle and Nuclear Physics\/}, {\em 62\/}(1), 102
  -- 134.
\newline\urlprefix\url{http://www.sciencedirect.com/science/article/pii/S0146641008000720}

\bibitem[{Burgess et~al.(1994)Burgess, Godfrey, K{\"o}nig, London, \&
  Maksymyk}]{burgess1994model}
Burgess, C., Godfrey, S., K{\"o}nig, H., London, D., \& Maksymyk, I. (1994).
\newblock Model-independent global constraints on new physics.
\newblock {\em Physical Review D\/}, {\em 49\/}(11), 6115.

\bibitem[{Carroll(2017)}]{carroll2017big}
Carroll, S.~M. (2017).
\newblock {\em The big picture: on the origins of life, meaning, and the
  universe itself\/}.
\newblock Penguin.

\bibitem[{Chalmers \& McQueen(2014)}]{Chalmers2014ConsciousnessAT}
Chalmers, D., \& McQueen, K. (2014).
\newblock Consciousness and the collapse of the wave function.
\newblock In {\em Quantum Mechanics and Consciousness\/}.

\bibitem[{Fox et~al.(2012)Fox, Harnik, Kopp, \& Tsai}]{Fox_2012}
Fox, P.~J., Harnik, R., Kopp, J., \& Tsai, Y. (2012).
\newblock Missing energy signatures of dark matter at the lhc.
\newblock {\em Physical Review D\/}, {\em 85\/}(5).
\newline\urlprefix\url{http://dx.doi.org/10.1103/PhysRevD.85.056011}

\bibitem[{Manohar(2020)}]{Manohar:2018aog}
Manohar, A.~V. (2020).
\newblock {Introduction to Effective Field Theories}.
\newblock {\em Les Houches Lect. Notes\/}, {\em 108\/}.

\bibitem[{Mele(2015)}]{mele2015measurement}
Mele, S. (2015).
\newblock The measurement of the number of light neutrino species at lep.
\newblock In {\em 60 Years of CERN Experiments and Discoveries\/}, (pp.
  89--106). World Scientific.

\bibitem[{Milgrom(1983)}]{milgrom1983modification}
Milgrom, M. (1983).
\newblock A modification of the newtonian dynamics as a possible alternative to
  the hidden mass hypothesis.
\newblock {\em The Astrophysical Journal\/}, {\em 270\/}, 365--370.

\bibitem[{Penrose(1989)}]{penrose1989emperor}
Penrose, R. (1989).
\newblock {\em The Emperor's New Mind: Concerning Computers, Minds, and the
  Laws of Physics\/}.
\newblock Oxford University Press.

\bibitem[{Peskin \& Schroeder(2015)}]{peskin2015introduction}
Peskin, M.~E., \& Schroeder, D.~V. (2015).
\newblock {\em An Introduction to Quantum Field Theory\/}.
\newblock CRC press.

\bibitem[{Polchinski(1984)}]{Polchinski:1983gv}
Polchinski, J. (1984).
\newblock {Renormalization and Effective Lagrangians}.
\newblock {\em Nucl. Phys. B\/}, {\em 231\/}, 269--295.

\bibitem[{Pullman(2000)}]{pullman2000}
Pullman, P. (2000).
\newblock {\em The Amber Spyglass\/}.
\newblock Scholastic/David Fickling Books.

\bibitem[{Rivat \& Grinbaum(2020)}]{rivat2020philosophical}
Rivat, S., \& Grinbaum, A. (2020).
\newblock Philosophical foundations of effective field theories.
\newblock {\em The European Physical Journal A\/}, {\em 56\/}(3), 1--10.

\bibitem[{Weinberg(1995)}]{weinberg1995quantum}
Weinberg, S. (1995).
\newblock {\em The Quantum Theory of Fields\/}, vol.~1.
\newblock Cambridge University Press.

\bibitem[{Weinberg(1996)}]{Weinberg:1996kw}
Weinberg, S. (1996).
\newblock {What is quantum field theory, and what did we think it is?}
\newblock In {\em {Conference on Historical Examination and Philosophical
  Reflections on the Foundations of Quantum Field Theory}\/}, (pp. 241--251).

\bibitem[{Wilczek(2015)}]{wilczek2015beautiful}
Wilczek, F. (2015).
\newblock {\em A Beautiful Question: Finding Nature's Deep Design\/}.
\newblock Penguin.

\bibitem[{Wilson \& Kogut(1974)}]{Wilson:1973jj}
Wilson, K., \& Kogut, J.~B. (1974).
\newblock {The Renormalization group and the epsilon expansion}.
\newblock {\em Phys. Rept.\/}, {\em 12\/}, 75--199.

\bibitem[{Wilson(1971{\natexlab{a}})}]{Wilson:1971bg}
Wilson, K.~G. (1971{\natexlab{a}}).
\newblock {Renormalization group and critical phenomena. 1. Renormalization
  group and the Kadanoff scaling picture}.
\newblock {\em Phys. Rev. B\/}, {\em 4\/}, 3174--3183.

\bibitem[{Wilson(1971{\natexlab{b}})}]{Wilson:1971dh}
Wilson, K.~G. (1971{\natexlab{b}}).
\newblock {Renormalization group and critical phenomena. 2. Phase space cell
  analysis of critical behavior}.
\newblock {\em Phys. Rev. B\/}, {\em 4\/}, 3184--3205.

\end{thebibliography}


\end{document}